\documentclass[aps,prb,reprint,twocolumn,superscriptaddress]{revtex4-2}

\usepackage{graphicx}
\usepackage{bm}
\usepackage{amsmath}  
\usepackage{amssymb}
\usepackage{comment}
\usepackage{hyperref}
\usepackage{xcolor}
\usepackage{relsize}
\usepackage{array}
\usepackage{multirow}
\usepackage{physics}
\usepackage{lipsum}
\usepackage[T1]{fontenc}
\usepackage{booktabs}
\usepackage{appendix}
\usepackage{multirow}

\def\qbs{\bm{q}}
\begin{document}
\title{Importance of electron-phonon coupling near the 
electron-liquid to Wigner-crystal transition in two-dimensional atomically thin materials} 

\author{Tixuan Tan}
\thanks{These authors contributed equally to this work.}
\affiliation{Department of Physics, Stanford University, Stanford, CA 94305, USA}
\author{Vladimir Calvera}
\thanks{These authors contributed equally to this work.}
\affiliation{Department of Physics, Stanford University, Stanford, CA 94305, USA}
\affiliation{Kavli Institute for Theoretical Physics, University of California, Santa Barbara, CA 93106, USA}
\author{Steven A. Kivelson}
\email{kivelson@stanford.edu}
\affiliation{Department of Physics, Stanford University, Stanford, CA 94305, USA}

\begin{abstract}
We study the effect of electron-phonon coupling on the location of the Fermi Liquid to Wigner Crystal transition in the two-dimensional electron gas realized in various material platforms. Based on dimensional estimates of the relevant parameters, we conclude that (as conventionally assumed) phonons are negligible in traditional semiconductor quantum well systems, but likely play a significant role in various recently synthesized atomically thin two-dimensional materials.
\end{abstract}
\maketitle

\section{Introduction}

The Fermi Liquid (FL) to Wigner Crystal (WC) transition 
reflects the the competition between the kinetic 
and Coulomb interaction energies in the two-dimensional electron gas (2DEG). The Coulomb interaction dominates 
at large average inter-electron 
spacing
when the electrons spontaneously crystallize into a WC. Since the initial proposal of its existence 80 years ago (\cite{PhysRev.46.1002}), enormous efforts, numerical/ theoretical \cite{PhysRevLett.102.126402,PhysRevB.39.5005,PhysRevB.18.3126,PhysRevLett.45.566,PhysRevLett.88.256601,PhysRevLett.42.798}, and experimental \cite{PhysRevLett.65.2189,Jang_2016,PhysRevLett.42.795,PhysRevLett.60.2765,PhysRevB.38.7881,PhysRevLett.66.3285,PhysRevB.40.4179,PhysRevB.46.7957,PhysRevLett.89.176802,PhysRevLett.91.016801,Chen_2006,tiemann2014nmr} 
, have been devoted to pin down the exact properties of the WC phase and the transition as a function of the 2D electron density $n_e$.

Even weak quenched disorder is known to destroy long-range WC order in 2D, 
but if the disorder is weak enough, the physics of the ideal clean limit can still be relevant. Recent progress in experimental techniques has made atomically thin two-dimensional materials of increasingly high mobility available that presumably mean low disorder. This is consistent with multiple reports of realization of WC phases in these platforms (either pristine or moir\'e)\cite{li2021imaging,sung2023observationelectronicmicroemulsionphase,Smole_ski_2021,zhou2021bilayer,regan2020mott,tsui2024direct}. Theoretical studies specifically targeted at these new atomically thin platforms include \cite{PhysRevB.95.115438,matty2022melting,PhysRevB.103.125146}

The FL to WC transition 
has typically been treated as a purely electronic problem. 
The underlying ion lattice where the electron gas lives is 
thus neglected.
In this paper, we 
address the importance of electron-phonon coupling in determining the FL-WC phase transition. 
In particular, 
in the adiabatic limit (i.e. when the ratio of the typical ionic mass to the electron effective mass, $M/m^\star\to \infty$)
we derive an expression for the 
difference in the energy per 
electron in the crystal phase relative to the fluid phase:
\begin{equation}\label{eq:energyShift}
\delta E_{WC}= - \sum_{\bm{G}\neq \bm{0}}A(\bm{G})\langle \rho_{\bm{G}} \rangle^2
\end{equation}
where $\bm{G}$ is summed over the reciprocal lattice vectors of the WC, $\langle\rho_{\bm{G}}\rangle$ is the expectation value of the $\bm{G}$ component of the electron density operator $\rho_{\bm{G}}$ in the WC state, and $A(\bm{Q})\geq 0$ can be expressed solely in terms of the electron-phonon coupling constants and the phonon elastic matrix. (cf. Appendix \ref{app:pathintegral}.)

To leading order in the electron-lattice coupling, $\langle \rho_{\bm{G}}\rangle$ can be computed neglecting this coupling;  we estimate it by treating the electronic motions in the WC state in the harmonic approximation, which is valid at low enough electron density $n_e$, i.e. when the Wigner-Seitz radius, $r_s \equiv (a_B^*\sqrt{\pi n_e})^{-1}$,  is sufficiently large.  
(Here, $a^*_B=\frac{4\pi\epsilon_0 \epsilon}{e^2 m^*}$ is the effective Bohr radius where $\epsilon$ is the background dielectric constant.)
For the relevant range of $r_s$, we find that $\langle \rho_{\bm{G}} \rangle$  is sufficiently strongly suppressed by the Debye-Waller factor at large $|\bm{G}|$ 
that it is a good approximation to keep only the first shell in the summation, $|\bm{G}|=\abs{\bm{G}_1} =
\frac{2\pi\sqrt{2n_e}}{\sqrt[4]{3}}$:  
\begin{equation}
\delta E_{WC}\approx -6A(\bm{G}_1)\langle \rho_{\bm{G}_1} \rangle^2 \approx - \frac{6E_0}N\exp(-\frac{C}{\sqrt{r_s}}),
\label{deltaE}
\end{equation}
where $N$ is the number of 
 unit cells of the host crystal per electron, 
$E_0$ is an overall 
energy scale that is dependent on the electron-phonon coupling strength, but independent of the electron density, $6$ is the number of reciprocal vectors $|\bm{G}|=|{\bf G}_1|$, and   
 $C \approx 12.9 $ is calculated in Sec.\ref{sec:change}.
The exact 
expression for 
$E_0$ is dependent on the details of the underlying crystal structure.
However, we obtained an estimate of its magnitude based on dimensional analysis, yielding $E_0 \sim \text{Ry}^* \equiv (1/2m^*)(\hbar/a_B^*)^2$ 
(with a proportionality constant that depends on the number and character of distinct phonon branches).

Since in all cases of interest, $N\gg 1$ (necessary for the effective mass approximation to be applicable), the effects of electron-phonon coupling are   intrinsically small.  However, state of the art numerics reveal that   the difference in energy between the FL and the WC is surprisingly small (of order 0.01\% $\text{Ry}^*.$ \cite{PhysRevLett.102.126402,azadi2024quantummontecarlostudy,smith2024groundstatephasestwodimension}) over a range of $r_s$ near critical value, $r_{c} \sim 30$.  Thus, even a relatively small energy can produce a significant change in $r_{c}$.


In a quantum well system, $N=W /(n_ea^3)$, where $W$ is the width of the quantum well and $a^3$ is the volume of a crystalline unit cell. 
For a typical GaAs 
heterostructure with $W \sim a_B^*\sim 10 a$,
$N \sim (\pi\times 10^3) r_s^2 $,
meaning that, at least from a thermodynamic persepctive, electron-phonon coupling is negligible.  However, for atomically thin 2D materials, $ N= 1/(n_ea^2) $, where $a^2$ is the area of the crystalline unit cell. For example, in monolayer $\text{WSe}_2$\cite{shih2023wse2}), $a_B^*\sim 3a$ so $N \sim 25 r_s^2$ while in bilayer $\text{MoSe}_2$\cite{xiang2024wigner} where melting of Wigner Crystal has been observed, $N\sim 200$, meaning that in both cases the effect of phonons on the FL-WC transition is likely non-negligible.

\section{Change in Energy associated with Electron-Phonon Coupling}\label{sec:change}

The  Hamiltonian describing the electron-phonon coupling is
\begin{equation}
H_{e\text{-}p}
=\frac{1}{\sqrt{N_i}}\sum_{r,\bm{k},\bm{q},\sigma}
\gamma_{\bm{q},r}
X_{\bm{q},r}c^{\dagger}_{\bm{k}+\bm{q},\sigma}c_{\bm{k},\sigma}\\
\end{equation}
where $X_{\bm{q,r}}=\frac{1}{\sqrt{2 M \Omega_{\bm{q},r}}}(b_{\bm{q},r}+b^{\dagger}_{-\bm{q},r})$ is the phonon coordinates, $N_i$ is the number of ion lattice unit cells, $r$ labels the phonon branches, $\sigma$ 
the spin of the electron, $\bm{k}$ and $\bm{q}$ are momentum labels, $b^{\dagger}\ \& \ b$ and $c^{\dagger}\ \& \ c$ are, respectively, the  phonon and electron creation \ \& annihilation operators, $M$ is the ion mass,
\footnote{More generally, with more than one species of atom per unit cell, $M$ is an appropriate average of the ion masses, which can in general depend on $r$ and $\qbs$.}
and $\Omega_{\bm{q},r}$ are the phonon frequencies. 
The electron-phonon coupling, $\gamma$,  can be estimated as an electronic energy scale over a lattice constant $\gamma \sim \text{Ry}^*/a$.

 We employ the Born-Oppenheimer 
 approximation, in which the electrons are in an electronic ground-state (either corresponding to a FL or a WC) 
 in the presence of a given ionic configuration.  The ground-state ionic configuration is then determined by minimizing the resulting adiabatic energy. 
The resulting change in the energy per electron to second-order in the electron-phonon coupling is (cf. Appendix \ref{app:pathintegral}) 
\begin{equation}
 \delta E=-\sum_{\bm{q},r}\frac{|\gamma_{\bm{q},r}|^2}{2M\Omega^2_{\bm{q}r}N_eN_i} \langle \rho_{\bm{q}} \rangle^2
\end{equation}
where $\langle \rho_{\bm{q}}\rangle =\sum_{j}\langle \exp(i\bm{q}\cdot \bm{r}_j)\rangle$, with $j$ labelling electrons, $\Omega_{\bm{q}r}$ is the energy of the phonon.

 It follows that electron-phonon coupling 
 lowers the energy of the WC  
 more than that of the FL, since for the later  $\langle \rho_{\bm{q}} \rangle$ vanishes when $\bm{q}\neq 0$, while the 
 $\bm{q}=0$ contribution for the two phases are the same at a fixed density. 

We estimate $\langle \rho_{\bm{q}} \rangle$ 
in the WC state using the harmonic approximation to treat quantum effects.
Defining $\bm{u}$ as the displacement of the  electron from its equilibrium position in the WC phase, we obtain

\begin{equation}
\langle \rho_{\bm{q}} \rangle
=N_e \sum_{\bm{G}} \delta_{\bm{q},\bm{G}}\ \exp(-\frac{ |\bm{q}|^2\langle  |\bm{u} |^2\rangle}{4})
\end{equation}
$\bm{G}$ is summed over the reciprocal lattice of the WC, and $\langle \rho_{\bm{q}} \rangle$ is attenuated by the Debye–Waller factor
(cf. Appendix~\ref{app:Debye-Waller Factor}). We expand $\bm{u}$ in terms of the WC phonon creation/annhilation operators $a^{\dagger}/a$: 
\begin{equation}
\bm{u}=\frac{1}{\sqrt{N_e m^*}}\sum_{\lambda=1}^2\sum_{\bm{p} \in BZ}\sqrt{\frac{1}{2\omega_{\bm{p},\lambda}}}(a^{\dagger}_{-\bm{p}\lambda}+a_{\bm{p}\lambda})\bm{\epsilon}^{\lambda}(\bm{p})
\end{equation}
where  $\lambda$ labels the normal modes of the WC, $\bm{\epsilon}^\lambda$ is the corresponding polarization vector, $\omega_{\bm{p},\lambda}$ is the energy of WC ``electronic phonon'', $\bm{p}$ is summed over the Brillouin zone (BZ) of the WC. The variance of $\bm{u}$ is thus expressed as 
\begin{equation}
\langle |\bm{u}|^2\rangle=\frac{1}{2 m^* A_{BZ}} \sum_{\lambda}\int_{\bm{p}\in BZ}\frac{1}{\omega_{\bm{p},\lambda}} d\bm{p},
\end{equation}
where $A_{BZ}$ is the area of WC Brillouin zone. The electronic phonon dispersion of the WC was calculated in Ref.\cite{PhysRevB.15.1959}. Using results therein, we obtain
\begin{equation}
\begin{split}
    \langle |\bm{u}|^2\rangle &= \frac{\sqrt{3}
    \tilde C}{16\omega_0 m^*}\\
    \tilde C&\equiv \sum_{\lambda}\int_{BZ}\frac{\omega_0}{\omega_{\bm{p},r}}  (\frac{a_0^2\dd^2\bm{p}}{\pi^2})\approx 17.693 \\
    \omega^2_0&=\frac{2e^2}{\pi\epsilon_0\epsilon_r m^* a_0^3}=\frac{8}{m^{*2} a_0^3 a^*_B}
\end{split}
\end{equation}
where $a_0 = \sqrt{\frac{2}{\sqrt{3}n_e}}$ is the lattice constant of the WC, $\omega_0$ is the typical energy scale of the WC electronic phonon.

Putting together the results above, we obtain an expression for the change in energy per electron in the WC phase (compared with the FL phase),
\begin{equation}
\delta E_{WC}=-\sum_{\bm{G}\neq 0,r}\frac{ |\gamma_{\bm{G},r}|^2}{2 N M\Omega^2_{\bm{G}r}} \exp(-\left|\frac{\bm{G}}{\bm{G}_1}\right|^2\frac{C}{\sqrt{r_s}})
\end{equation}
where we defined the constant $C=
\tilde C\frac{3^{5/8} \pi ^{7/4}}{12\ 2^{3/4}}\approx 12.9$. Numerically and experimentally, the reported critical $r_s$ for FL-WC transition is around 30 \cite{PhysRevLett.102.126402,sung2023observationelectronicmicroemulsionphase}.   We have verified that for $r_s \in [10,50]$, 
the first shell dominates in the summation $\sum_{\bm{G}}$. Therefore, as an estimate, we can keep only the terms corresponding  to the first shell, resulting in the expression in Eq. \ref{deltaE} with
 \begin{equation}
    E_0 \equiv \frac{1}{6}\sum_{r,\bm{G}; \abs{\bm{G}}=\bm{G}_1}\frac{|\gamma_{\bm{G},r}|^2} {2M\Omega^2_{\bm{G},r}}\nonumber.
\end{equation}

The exact form of $\gamma_{\bm{q},r}$ is highly 
material-specific, 
in ways we will not explore here.  Instead, we notice that $E_0$ has dimensions of energy, and is independent of the ion mass since $\Omega^2\sim M^{-1}$.
Thus, we expect generically that  $E_0 \sim \text{Ry}^*$.

\begin{table*}[t]
\begin{tabular}{ |p{3cm}||p{3cm}|p{3cm}|p{3cm}|p{3cm}|  }
 \hline
 Transition & 
$N_0 \to \infty$
& $N_0$=$1\times 10^4$ &$N_0$=$2\times 10^3$ &$N_0$=$1\times 10^3$ \\
 \hline
 \hline
 PM FL to FM WC   & $1
 \times 10^{-3}\ \ (
 r_c=33)$    &$1
 \times10^{-3}\ \ (
 r_c=26)$&   $3
 \times 10^{-3}\ \ (
 r_c=16)$&$5
 \times 10^{-3}\ \ (
 r_c=13)$\\
PM FL to AF WC& $8
\times 10^{-4}\ \ (
r_c=31)$   & $1
\times 10^{-3}\ \ (
r_c=24)$  &$3
\times 10^{-3}\ \  (
r_c=15)$ &$6
\times 10^{-3}\ \ (
r_c=12)$\\ 
 \hline
\end{tabular}
\caption{Estimate of the 
extent of the microemulsion phase, $\delta n/n_c$, taking $E_0=1\text{Ry}^*$ based on the expressions in Eq.\ref{eq_microemulsion} and Eq.\ref{eq:finalenergy} and the numerical results for the purely electronic problem (first column - $N_0\to \infty$) 
from Ref.\cite{PhysRevB.108.L241110}. The critical value of $r_s=r_c$ is determined by combining the fitting formula in Ref.\cite{PhysRevLett.102.126402} and Eq.\ref{eq:finalenergy}. AF/FM stands for antiferromagnetic/ferromagnetic.
}
\label{tb_modifiedn}
\end{table*}
\section{Implications}

\subsection{Shift of the critical $r_s$ for FL-WC transition}

\begin{figure}
\begin{center}
    \includegraphics[width=0.48\textwidth]{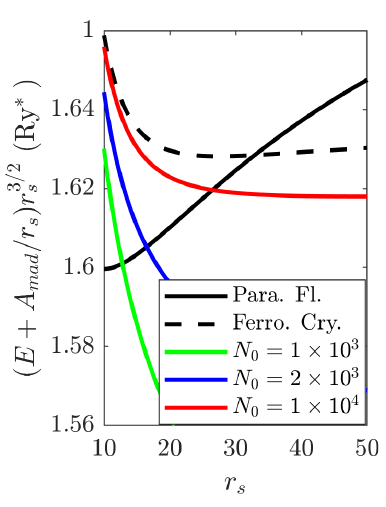}
    \caption{The calculated energy per particle $E$ as a function of $r_s$ of the paramagnetic fluid  and ferromagnetic crystalline  from the variational MC results of Ref.\cite{PhysRevLett.102.126402,rapisarda1996diffusion} are shown as the solid and dashed black line, respectively.  The energy of the WC shifted due to the  
    electron-phonon coupling 
    is represented by the solid colored lines for various values of $N_0$. 
    Results there are plotted in atomic units. $A_{mad}=2.12206 \text{Ry}^*$ is associated with the Madelung energy of static crystal \cite{PhysRevLett.102.126402}.
    }
    \label{fig_modifiedenergy}
\end{center}
\end{figure}

The density dependence (or in other words the  $r_s$ dependence) of $\delta E$ enters through the explicit dependence of the Debye-Waller factor and through the implicit dependence from
$N\propto r_s^2$.  To make this apparent, 
we express the difference in energy between the WC and the FL as
\begin{equation}
\delta E_{WC}=  -6\frac{E_0}{N_0}\left(\frac{r_0}{r_s}\right)^2 \exp(-\frac{C}{\sqrt{r_s}})
\label{eq:finalenergy}
\end{equation}
where $r_0$ is a reference value of $r_s$,
and $N_0$ is $N$ evaluated at $r_s =r_0$. For convenience, we take $r_0=30$, characteristic of the expected value at which the FL-WC transition occurs. $N_0$ can be obtained for specific meaterials directly from the measured electron density in experiments. As already mentioned, values for  bilayer 
and monolayer MoSe$_2$ can be estimated to be  $N_0\sim 200$ and $N_0\sim 
 2000$, respectively
\cite{xiang2024wigner,sung2023observationelectronicmicroemulsionphase,Smole_ski_2021}.
In Fig.\ref{fig_modifiedenergy}, starting from values of the energy of the FL and WC states in the absence of elecron-phonon coupling taken from state of the art variational Monte-Carlo calculations\cite{PhysRevLett.102.126402}, we show the shift in the relative energy that would result from $N_0$ in this range, and assuming a value of $E_0=\text{Ry}^*$. 
Clearly, the change in the critical $r_s$ is significant. For instance, for $N_0=1000$, the critical $r_s$ for the transition from a ferromagnetic WC to a paramagnetic FL is shifted from $r_s=32.89$\cite{PhysRevLett.102.126402} to $r_s=12.78$! 

\subsection{The extent of microemulsion phases}

In principle, macroscopic phase separation is precluded in the presence of long-range Coulomb interactions. Rather than a direct first order transition between the FL and the WC, there should arise an intermediate phase or phases  - most likely microemulsion phases \cite{spivak2006transport,PhysRevB.70.155114}. However, it was pointed out in Ref.\cite{PhysRevB.108.L241110} that for the pure 2DEG, this intermediate phase is expected to be extremely narrow, due to the fact that the difference in the chemical potential between the two extremal phases, $\delta \mu$, appears (from the numerics) to be so small.  Specifically, in the same paper, an approximate upper bound on the range of density, $\delta n$, of the microemulsion phases was found to be
\begin{equation}
\label{eq_microemulsion}
\frac{\delta n}{n_c}\lesssim 
 \frac{3\sqrt{\pi}}{16}\ r_s \ \frac{\delta \mu}{\text{Ry}^*}
\end{equation}
where $n_c$ is the critical electron density at the FL-WC transition.

Notably, as is apparent from Fig.~1, the electron-phonon effects can substantially increase $\delta \mu$.  In Table~\ref{tb_modifiedn} we show  
the corresponding 
estimates of $\frac{\delta n}{n_c}$ obtained from the above formula for various assumed values of $E_0/N_0$. Even for relatively large $N_0$, there is a significant enhancemnt, albeit not large enough to account for the range of microemulsion phases reported in experiment, e.g. $\frac{\delta n}{n_c}\sim O(1)$ in Ref. \cite{sung2023observationelectronicmicroemulsionphase}.


\section{Discussion
}

In this article, we have 
carried through an initial (perturbative) study of the importance of electron-phonon coupling 
on the thermodynamic properties of the 2DEG proximate to the FL to WC transition.  For circumstances characteristic of the 2DEG in semiconductor quantum wells, we found that because the typical electronic wave-functions are spread out over so many unit cells of the underlying crystal, the effects of electron-phonon coupling are negligible.  However, for parameters characteristic of novel 2D materials, at the very least the critical value of $r_s$ is expected to be significantly shifted (to smaller values) by the coupling to phonons.   
We also find a corresponding phonon-induced enhancement of the stability of intermediate microemulsion phases, although it is unclear whether this effect can be large enough to account for recent reported observations of these phases.

There are several further refinements of the theory presented here that may be worth pursuing.  In the first place, when the shift of the critical $r_s$ is substantial, the approximations we have employed may no longer be reliable - both treating the WC structure factor in harmonic approximation (which is   less well justified for smaller $r_s$) and treating the electron-phonon coupling perturbativey.  Secondly, the adiabatic approximation, valid when the phonon frequencies $\Omega_{\bm {q},r}$ are small compared to the characteristic electron frequencies, $\sim \text{Ry}^*/\sqrt{r_s}$, may breakdown in some circumstances.  In Appendix \ref{app:pathintegral} we obtain a formal expression for $\delta E$ to second order in the electron-phonon coupling 
including the effects of dynamical phonon ground-state fluctuations.  However, we have not systematically explored the effect of these terms.  Finally, there may be circumstances in which our crude dimensional estimate of the strength of the electron-phonon coupling may underestimate the effect of particularly strongly coupled phonon modes.

Finally, to be directly relevant to experiment, the effects of finite temperature and of effects of the underlying lattice beyond the effective mass approximation (including the possibility of commensurate locking to the underlying lattice) need to be considered.

\noindent{\bf Acknowledgements:}
 We acknowledge useful conversations with Z.Y. Han, Z.Y. Zhu., I. Esterlis. 
 This work was supported in part by NSF-BSF award DMR2310312 at Stanford (V.C. and S.A.K.). V.C. was also supported in part by grant NSF PHY-2309135 to the Kavli Institute for Theoretical Physics (KITP).
 T.T. is supported by Stanford Graduate Fellowship.

\newpage
\appendix
\section{Deybe-Waller Factor}\label{app:Debye-Waller Factor}
For a system of harmonic oscillators, it is proved in Ref.\cite{david1966short} that for operators $A$ that is linear combinations of ladder operators
\begin{align}
A=\sum_{i}c_i a^{\dagger}_i+ d_i a_i\\
\langle \exp(A) \rangle=\exp(\frac{\langle A^2 \rangle}{2})
\end{align}
where $i$ labels the the harmonic oscillators, $a_i/a^{\dagger}_i$ is the corresponding ladder operators.
The expectation value of $\exp(A)$ is evaluated with respect to an ensemble of harmonic oscillators $\rho = \exp(- \sum_{i} \epsilon_{i}a^{\dagger}_i a_{i})$.

For us, the ground state of WC is the vacuum of the WC phonons under harmonic approximation. Thus,
\begin{equation}
\begin{split}
\langle \rho_{\bm{q}} \rangle&=\sum_i \langle \exp(i\bm{q}\cdot \bm{r}_i)\rangle\\
=& \sum_i \langle \exp(i\bm{q}\cdot \bm{u}_i)\rangle \exp(i\bm{q}\cdot \bm{R}_i)\\
=&\sum_i \exp(i\bm{q}\cdot \bm{R}_i)\exp(-\frac{\langle (\bm{q}\cdot \bm{u}_i)^2 \rangle}{2})\\
=&N_e (\sum_{\bm{G}} \delta_{\bm{q},\bm{G}}) \exp(-\frac{\langle (\bm{q}\cdot \bm{u})^2 \rangle}{2})
\end{split}
\end{equation}

where we defined $\bm{u}_i$ as the displacement of the i-th electron from its equilibrium position $\bm{R}_i$. We have used the translation invariance of the ground state $\langle (\bm{q}\cdot \bm{u}_i)^2 \rangle=\langle (\bm{q}\cdot \bm{u})^2 \rangle$. To proceed further, we exploit the rotational symmetry of WC. We define ($\alpha,\beta=x,y$)
\begin{equation}
U_{\alpha\beta}=\langle u_{j\alpha}u_{j,\beta} \rangle,
\end{equation}

By $C_3$ symmetry of the WC ground state, $U$ matrix obeys the following relation
\begin{equation}
    U_{\alpha\beta}=(C_3)_{\alpha\alpha'}(C_3)_{\beta\beta'}U_{\alpha'\beta'}
\end{equation}
where $(C_3)_{\alpha\beta}$ is the matrix element of the matrix representing $\frac{2\pi}{3}$ rotation. Since $U_{\alpha\beta}$ is a $2\times2$ real symmetric matrix, the above relation proves $U$ matrix is proportional to identity, and thus equal to $U_{\alpha\beta} = \frac{\delta_{\alpha\beta}}{2} \left(\sum_{\gamma=1}^{2}U_{\gamma\gamma}\right)$.
Thus, 
\begin{equation}
\begin{split}
\langle \rho_{\bm{q}} \rangle
=&N_e (\sum_{\bm{G}} \delta_{\bm{q},\bm{G}}) \exp(-\frac{ |\bm{q}|^2\langle  |\bm{u} |^2\rangle}{4}).
\end{split}
\end{equation}

\section{Graphene phonon}

Graphene is one of the earliest atomically thin 2D materials. Electronic properties and phonon properties of it has been well modelled. Thus, as an example, we study the electron-phonon coupling in graphene in this appendix as a verification of the ideas we exploited in the main text.

Graphene tight-binding Hamiltonian with nearest neighbour hopping  can be written as
\begin{equation}
\hat{H}=-\sum_{m,i,s}t_{s,i}c^{\dagger}_{m,s}c_{m(i),s(i)}
\end{equation}
where $s=A,B$ labels the sublattice, and $m$ labels the unit cell, $i=1,2,3$ goes over the three nearest neighbour pair. $c^{\dagger}/c$ are electronic creation annihilation operators, the positions for the $s$ sublattice atom in the $m$ unit cell is written as $\bm{R}_{m,s}=\bm{R}_{m}+\bm{R}_{s}$. The hopping $t$ depends on the distance between the two sites involved.
\begin{equation}
\begin{split}
    t_{s,i}(r)
    &=t(|\bm{R}_{m,s}-\bm{R}_{m(i),s(i)}|)\\
    &=t_0 \exp(-(r-a_0)/d_0),
\end{split}
\end{equation}
$d_0=0.452\text{\AA}$, $t_0=2.7\text{eV}$, 
$a_0=1.42\text{\AA}$ is the distance between nearest neighbour carbon atoms\cite{PhysRevLett.127.167001}. The displacement field can be written as
\begin{equation}
\begin{split}
\bm{u}^m_{s}=&\sum_{r}\sum_{\bm{q}\in BZ}
\frac{1}{\sqrt{N_i M_s}}\sqrt{\frac{\hbar}{2\Omega_{\bm{q},r}}}(b_{\bm{q},r}+b^{\dagger}_{-\bm{q},r})\\
&\bm{\epsilon}^r_{s}(\bm{q})\exp(i\bm{q}\cdot \bm{R}_m)
\end{split}
\end{equation}
where $\Omega_{\bm{q},r}$ is the phonon energy of the graphene in the $r$-th branch, $N_i$ is the number of graphene unit cells, $M_s$ is the mass of carbon atom. $\bm{\epsilon}$ is the phonon polarization vector.
Then the electron-phonon Hamiltonian can be written as
\begin{equation}
\hat{H}_{e\text{-}ph}=-\sum_{m,i,s}(\bm{u}^{m}_s-\bm{u}^{m(i)}_{s(i)})\cdot \nabla t_{s,i}(r) c^{\dagger}_{m,s}c_{m(i),s(i)}
\end{equation}
We define the vectors pointing from each atom to its neighbours as $\bm{\delta}^s_i=\bm{R}_{m(i),s(i)}-\bm{R}_{m,s}$.

In terms of the ionic phonon operators
\begin{widetext}
\begin{equation}
\begin{split}
\hat{H}_{e\text{-}ph}=&-\sum_{i,s,r}\sum_{\bm{q},\bm{k},\bm{k'}\in BZ}\sum_{\bm{G}\in RL}
\frac{1}{\sqrt{N_i M_s}}
\sqrt{\frac{\hbar}{2\Omega_{\bm{q},r}}}(\bm{\epsilon}^{r}_{s}(\bm{q})e^{i\bm{k'}\cdot (\bm{\delta^s_i}+\bm{R}_s-\bm{R}_{s(i)})}-\bm{\epsilon}^{r}_{s(i)}(\bm{q})e^{i(\bm{k'}+\bm{q})\cdot (\bm{\delta^s_i}+\bm{R}_s-\bm{R}_{s(i)})})\cdot \nabla t_{s,i}(r) \\
&(b_{\bm{qr}}+b^{\dagger}_{-\bm{q}r})c^{\dagger}_{\bm{k},s}c_{\bm{k'},s(i)}\delta_{\bm{k'}-\bm{k}+\bm{q},\bm{G}}
\\
&\equiv \frac{1}{\sqrt{N_i}}\sum_{\bm{k},\bm{q},s,r} \gamma^{r,s}_{\bm{k},\bm{q}} X_{\bm{q},r}c^{\dagger}_{\bm{\lceil k+q \rceil },s}c_{\bm{k},\bar{s}}
\end{split}
\end{equation}
\end{widetext}
where we have used the fact that nearest neighbor pairs in graphene are in different sublattice $s$ and $\bar{s}$. The $\lceil .\rceil$ operator sends the vector back to the BZ. The coupling  is 
\begin{equation}
\begin{split}
\gamma^{r,s}_{\bm{k},\bm{q}}&=-\sum_{i}\nabla t_{s,i}(r)\cdot\\
&(\bm{\epsilon}^{r}_{s}(\bm{q})e^{i\bm{k}\cdot (\bm{\delta^s_i}+\bm{R}_s-\bm{R}_{s(i)})}-\bm{\epsilon}^{r}_{s(i)}(\bm{q})e^{i(\bm{k}+\bm{q})\cdot (\bm{\delta^s_i}+\bm{R}_s-\bm{R}_{s(i)})})
\end{split}
\end{equation}
For graphene, we are primarily interested in the electrons confined at the low-energy sector, i.e. the $\bm{K}/-\bm{K}$ valley. As such, we can drop the $\bm{k}$ dependence of the matrix elements, and replace all $\bm{k}$ in its definition by $\bm{K}$. Phonon dispersion and polarization vector of monolayer graphene is well-known in the literature \cite{PhysRevLett.127.167001,Wirtz_2004}. Using results there in, we may calculate the following quantity as a function of WC density
\begin{equation}
E_{r,s}\equiv \sum_{|\bm{G}|=|\bm{G}_1|}\frac{N_e}{N_i}\frac{|\gamma^{r,s}_{\bm{K},\bm{G}}|^2}{2M\Omega^2_{\bm{G},r}}.
\end{equation}
To make sense of the magnitude of this quantity and compare with the estimate we made in the main text, it is better to consider bilayer graphene, whose dielectric constant is around 5, and $m_e\approx 0.03 m_0$. The effective Rydberg is hence 16 meV. We assume that the electron-phonon coupling will not be too much different in monolayer graphene and bilayer graphene, so we use $E_{r,s}$, which is calculated for monolayer graphene, as a proxy for the corresponding quantity in bilayer graphene.

\begin{figure}[h!]
    \centering
    \includegraphics[width=0.48\textwidth]{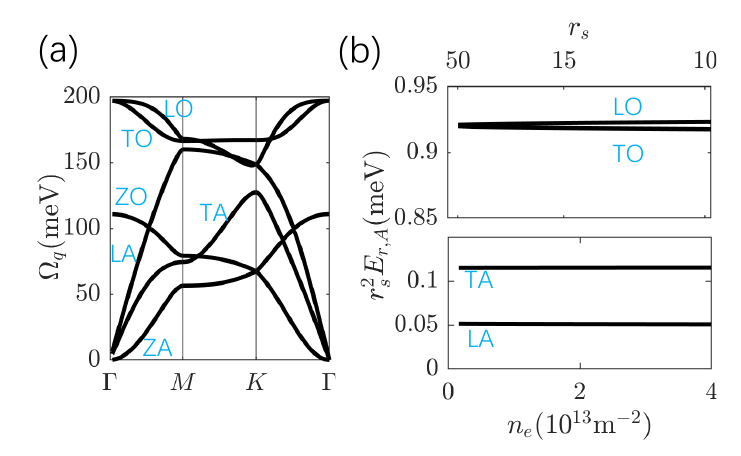}
    \caption{(a)Dispersion of Monolayer graphene phonon. T(transverse), L(longitudinal),Z(out-of-plane) labels the direction of polarization. A(Acoustic), O(Optical) labels the type of phonon. (b)electron-phonon coupling strength $E_{r,s}$ as a function of electron density $n_e$. $n_e$ is translated into $r_s$ by using the corresponding quantities of bernal-stacked bilayer graphene, $\epsilon_r=5$, $m^*=0.03m_e$.}
    \label{fig_phonondispersion}
\end{figure}

The calculation in the main text is done by taking $r_s^2\sum_{r,s}E_{r,s}\sim \frac{6r_0^2}{N_0} \text{Ry}^*\sim O(0.1) \text{Ry}^*$. As can be observed from Fig.\ref{fig_phonondispersion}, the calculated $\sum_{r,s}r_s^2E_{r,s}$ is indeed roughly this large, justifying the treatment in the main text\footnote{Technically, we should go to the band basis, but that's just a unitary transformation. That will not change the magnitude of the coupling strength. Since out goal here is to show the coupling strength magnitude is indeed roughly the value we estimated in the main text, we will not go to the band basis since that clutters the notation}.

\newpage
\section{Path integral 
formulation and the effect of non-infinite $M$.
}\label{app:pathintegral}

\def\qbs{\bm{q}}
We start with the path integral formulation of the electron-phonon problem 
\begin{equation}
    \begin{split}
        Z &= \int \mathcal{D} \Psi \int \mathcal{D} X e^{- S[\Psi,X]} \\
        S[\Psi,X] &= S_{e}[\Psi] +  S_{ph}[X] + S_{e\text{-}ph}[\Psi,X] \\
        S_{ph}[X] &= \int \dd{\tau}\sum_{\bm{q}} \left(\frac{M}{2}\abs{\dot{X}_{\bm{q}}}^2  + \frac{K_{\bm{q}}}{2}\abs{{X}_{\bm{q}}}^2\right)\\
        S_{e\text{-}ph}[\Psi, X] &= \int\dd{\tau} \sum_{\qbs} g_{\qbs}\rho_{\qbs} X_{\qbs}^*
    \end{split}
\end{equation}
where $\Psi$ and $X$ are the fermionic (electron) and bosonic (phonon) fields, $\rho$ is the density of the fermionic fields, and $g_{\qbs}$ is the coupling constant. For simplicity, we are considering a single phonon branch. Note that $X_{\qbs}^*=X_{-\qbs}$.

We next determine the corresponding saddle-point configurations:
\begin{align}
    K_{\qbs}\bar X_{\qbs}=-g_{\qbs}\bar \rho_{\qbs}
\end{align}
where $\bar\rho_{\qbs}$ is the expectation value of $\rho_{\qbs}$ in the ensemble defined by the saddle-point action, $S[\psi,\bar X]$.  In general, $\bar \rho$ is a functional of $\bar X$, and the above equation is thus a self-consistent equation of the sort familiar form Hartree-Fock theory.  However, in the present case, where the electron-phonon coupling is weak, $\bar \rho$ can be evaluated in the $g\to 0$ limit, in which case $\bar\rho_{\qbs}$ is zero for any non-zero $\qbs$ in any electron liquid phase and is non-zero for a set of reciprocal lattice vectors, $\bm{G}$, in the WC phase.  

We now define an unperturbed action, $S_0$, that characterizes the saddle-point ensemble, 
\begin{equation}
\begin{split}
 S_0[\Psi,\delta X]&\equiv S[\psi,\bar X]+S_{ph}[\delta X]\\
 &=S_e[\Psi]+\int d\tau \sum_{\bm{q}} g_{\bm{q}}\rho_{\bm{q}}\bar{X}_{-\bm{q}}\\
  &+\frac{M}{2}|\delta \dot{X}_{\bm{q}}|^2+\frac{K_{\bm{q}}}{2}|\delta X_{\bm{q}}|^2+\frac{K_{\bm{q}}}{2}|\bar{X}_{\bm{q}}|^2 
\end{split}
\end{equation}
and the remaining term, which taking advantage of the self-consistency equation above, can be written as {
\begin{align}
S[\Psi, X] &=S_0[\Psi,\delta X]+\delta S[\Psi,\delta X]\\
    \delta S[\Psi,\delta X]&= \int\dd{\tau} \sum_{\qbs} g_{\qbs}\ \delta\rho_{\qbs}\  \delta X_{\qbs}^*
\end{align}
where $\delta\rho_{\qbs}\equiv \rho_{\qbs}-\bar \rho_{\qbs}$.  

{The partition function then factorize into
\begin{align}
Z&=Z_0 \langle \exp(-\delta S) \rangle_0, \\
Z_0&=\int \mathcal{D} \Psi \int \mathcal{D} X e^{- S_0[\Psi,\delta X]}
\end{align}
and $\langle \dots \rangle_0$ is evaluated with respect to the action $S_0$. 

 Note that as we approach zero temperature ($T\to 0$), $Z\approx \exp(-E/T)$, where $E$ is the ground state energy. By the linked cluster theorem, for an operator $F$
\begin{equation}
\langle \exp(-F)\rangle_{0}=\exp(-\langle F \rangle_0+\frac{1}{2}\langle (F-\langle F \rangle_0)^2 \rangle_0... )
\end{equation}
We chose $\bar{{\rho}}_{\bm{q}}=\langle \rho_{\bm{q}} \rangle_0$ so that $\langle \delta S \rangle_0=0$. We can thus evaluate the changes in the ground-state energy perturbatively in powers of $\delta S$. To zeroth order in $\delta S$, we recover the results obtained using the approximations in the main text. The leading correction in $\delta S$ is second order because of our chose of $\bar{\rho}_{\bm{q}}$. Taking into account both corrections, we obtain
\begin{equation}
\begin{split}
    \Delta E & = -\frac{1}{2}\sum_{\qbs} |g_{\qbs}|^2\Big[\frac{|\langle \rho_{\qbs}\rangle_0|^2}{K_{\bm{q}}}  
    \\ &+ \int d\tau  \langle \delta X_{\qbs}^*(0)\delta X_{\qbs}(\tau)\rangle_0 \langle \delta \rho_{\qbs}^*(0)\delta \rho_{\qbs}(\tau)\rangle_0\Big]  
    \nonumber \\
    &+\dots \nonumber
    \end{split}
\end{equation}
where the first term is the saddle-point contribution discussed in the text, the second term is a fluctuation term, and $\dots$ refers to higher order terms in powers of $|g|^2$.  The fluctuation term (and all higher such terms) is expressed as an integral over products of purely electronic and purely lattice correlation functions.  We generally expect the characteristic relaxation rates of the lattice fluctuations to be slow compared to those of the electronic fluctuations, so to good approximation, which becomes exact in the $M\to\infty$ limit, we can replace the lattice correlators by the corresponding equal-time quantity - in this case $\langle |\delta X_{\qbs}|^2\rangle_0$. Better still, given that we have restricted our attention to the $T=0$ case, these are quantum fluctations, and hence vanish in proportion to $M^{-1/2}$. In the adiabatic approximation, these terms can be viewed as renormalizations of the phonon zero-point energies.

 Note that there are problems with this approach applied to the finite temperature case, where the acoustic mode fluctuations always give a divergent contribution to $\langle [\delta X]^2\rangle$.  We will not address this issue explicitly here, other than to note that while this is a significant feature of the physics - reflecting the fact that the finite $T$ WC has only quasi-long-range order, the key long-wave-length acoustic modes make very little contribution to the free energy and so are likely of negligible importance for present purposes.

\bibliography{biblio}

\end{document}